\documentstyle[11pt,newpasp,twoside,epsf]{article}
\begin{document}

\title{Double Bars, Inner Disks, and Nuclear Rings in Barred Galaxies}
\author{Peter Erwin}
\affil{Instituto de Astrof\'{\i}sica de Canarias\\
38200 La Laguna, Tenerife, Spain}
\author{Linda S. Sparke}
\affil{Astronomy Department, University of Wisconsin--Madison\\
475 N. Charter St.\\
Madison, WI 53706, USA}
\author{Juan Carlos Vega Beltr\'an, John Beckman}
\affil{Instituto de Astrof\'{\i}sica de Canarias}

\begin{abstract}
We present results of a high-resolution imaging survey of barred
S0--Sa galaxies which demonstrate that the central regions of these
galaxies are surprisingly complex.  We see many inner bars --- small,
secondary bars (typically less than a kiloparsec in radius) located
inside of, and probably rotating faster than, the large primary bars. 
These are present in about one quarter to one third of all our sample. 
In contrast to some theoretical expectations, they do \textit{not}
seem to enhance AGN activity significantly.  A third of barred S0's
appear to host kiloparsec-scale \textit{disks} within their bars; but
the frequency of such inner disks is much lower in our S0/a and Sa
galaxies.  In addition, we find one example of a \textit{triple}
barred galaxy, and two cases of purely \textit{stellar} nuclear rings
--- probably the fossil remnants of past circumnuclear starbursts.

We comment briefly on results from an ongoing analysis of known
double-barred systems, extending to Hubble types as late as Sbc, and
discuss their characteristic sizes and orientations.
\end{abstract}

\section{Introduction}

Double bars --- systems where a small, faster-rotating ``inner'' bar
forms concentrically within a large, ``outer'' bar --- were proposed
by Shlosman, Frank, \& Begelman (1989) as a mechanism for efficiently
feeding gas into the centers of galaxies and fuelling activity there. 
Though isolated examples of real double-barred galaxies were known
earlier (e.g., de Vaucouleurs 1974), high-resolution imaging has
revealed numerous examples in the last decade; see Friedli (1996) for
a review.  One observational signature is the twisting of isophotes
within a large bar; Elmegreen et al.\ (1996a) argued that such twists
were rather common in early-type barred galaxies.

Not all such isophote twists are due to inner bars, however: Erwin \&
\& Sparke (1999) showed that twists, and peaks in the ellipticity
profiles of isophotes, could also be caused by luminous, stellar
nuclear rings or by kiloparsec-scale \textit{disks} embedded within
bars.  Here, we discuss some results from a survey designed to find
out just how common inner bars --- and other structures inside bars
--- might be, and learn something of their characteristics.  A more 
complete description and analysis will appear in Erwin \& Sparke 
(2001a,b).

\section{Samples and Observations}

To minimize confusion due to dust, we concentrated on early type (S0
and Sa) galaxies.  We chose all barred (SB or SAB) S0 and Sa galaxies
in the UGC, north of $\delta = -10\deg$, which met the following
criteria: nearby ($z \le 2000$ km/s), large and easily resolved (major
axis $\ge 2\arcmin$), and low to moderate inclination (major : minor
axis ratio $\le 2$).  The Virgo Cluster was excluded, in order to
concentrate on field galaxies.  This resulted in a total of 38
galaxies (the ``WIYN sample''); twenty of these are S0, with ten S0/a
and eight Sa galaxies.  The diameter limit excludes some
low-surface-brightness galaxies, but this is probably the only
significant bias.

We observed all but two of the galaxies in $B$ and $R$ with the 3.5m
WIYN telescope in Tucson, Arizona, between December, 1995, and March,
1998.  Seeing ranged from 0.6--1\farcs3, with a median of 0\farcs8. 
We obtained archival HST images, both WFPC2 and NICMOS, for over half
the galaxies.  For NGC 936, we obtained $B$, $V$, and $R$ images
(0\farcs7) with the 4.2m William Herschel Telescope in December, 2000.

We have also begun compiling a catalog of known double-bar and
inner-disk systems, combining galaxies from our survey with other WIYN
observations and with galaxies drawn from the literature.  Restricting
ourselves to strong, unambiguous detections, we have taken information
from a number of studies, including Shaw et al.\ (1993); Wozniak et
al.\ (1995) and Friedli et al.\ (1996); Mulchaey, Regan, \& Kundu
(1997); Jungwiert, Combes, \& Axon (1997), M\'arquez et al.\ (1999);
and Greusard et al.\ (2000).  We use the published images and ellipse
fits --- as well as archival HST images wherever available --- to
remeasure bar position angles and sizes in a consistent manner, and to
eliminate mistaken inner bar/disk identifications due to nuclear
rings, strong dust lanes, etc.  We re-classify some galaxies as
inner-disk rather than double-bar, using criteria explained below.  To
date, we have measurements for a total of 23 double-bar and 17
inner-disk galaxies; this set is the ``expanded'' sample.

\section{Techniques and Identification}

To analyze the WIYN sample, we used a combination of ellipse fits to
isophotes, unsharp masking, and color maps to identify and measure a
variety of central structures in these barred galaxies: inner bars,
inner disks, nuclear rings, nuclear spirals, and near-nuclear
off-plane dust.  We were particularly careful to discriminate between
structures which can produce similar distortions in isophotes and
ellipse fits (Figure~1).

\begin{figure}
\plotone{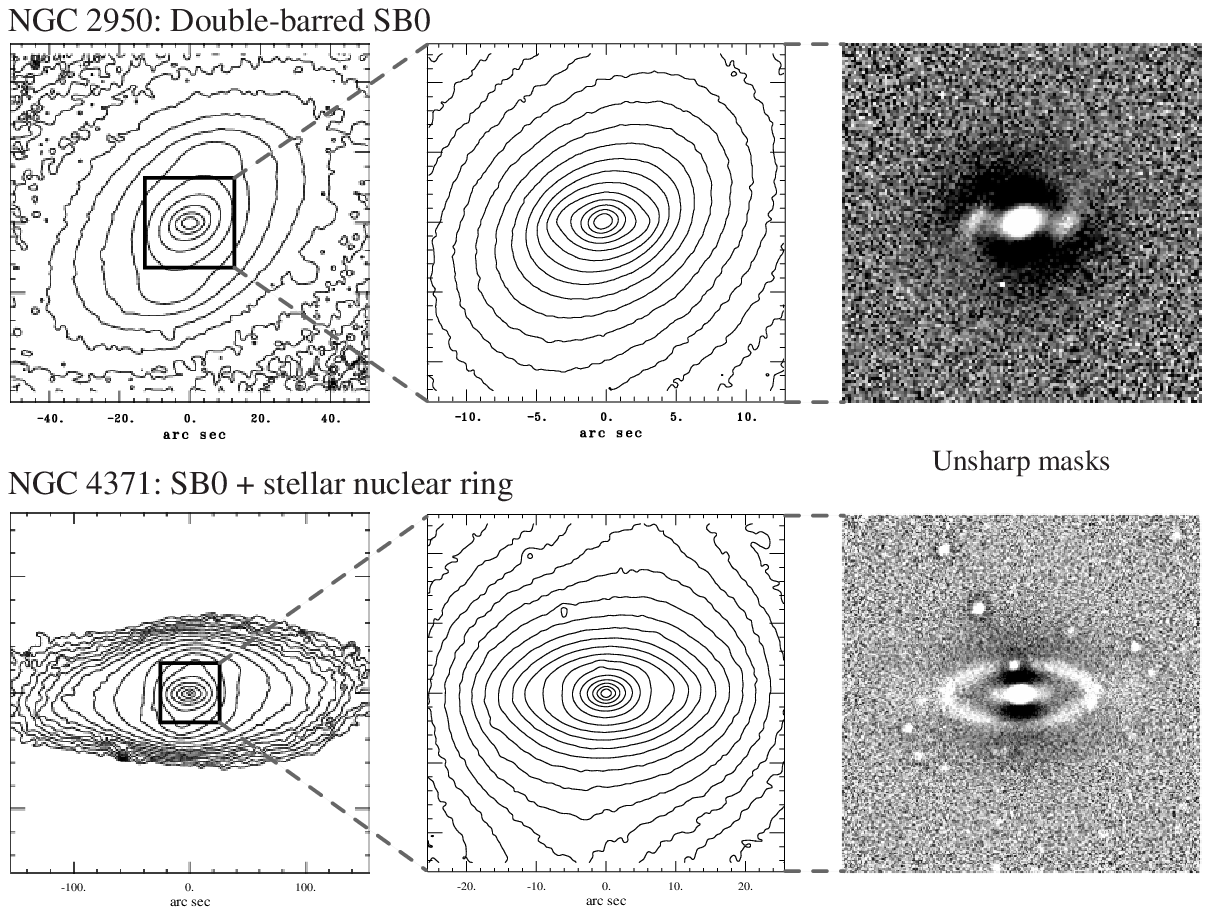} \caption{Above, the double-barred galaxy
NGC 2950; below, NGC 4371, a barred galaxy with a stellar nuclear ring
(WIYN $R$-band images for both).  In both cases, the inner structures
visibly distort the isophotes and create strong peaks in ellipse-fit
profiles; but the inner bar is distinctly different from the nuclear
ring in the unsharp masks (right-hand side).}

\end{figure}

\section{Results}

\subsection{Inner Bars}

Inner bars are surprisingly common: they occur in at least
one-quarter (possibly as many as 40\%) of the barred S0--Sa galaxies
in the WIYN sample, with no discernible dependence on Hubble type or
outer-bar strength.  One galaxy is actually \textit{triple}-barred
(NGC~2681; see Erwin \& Sparke 1999).  A typical inner bar in the WIYN
sample is 240--750 pc in radius (median = 400 pc), 6--14\% the size of
its ``parent'' outer bar (median = 12\%), and 3--8\% of its galaxy's
$R_{25}$ (median = 5\%).  In the expanded sample, we find a wider
range in relative and absolute sizes (240--1360 pc, 6--23\% of
outer-bar length, 3--13\% of $R_{25}$), but largely unchanged median
values (740 pc, 13\%, and 6\%, respectively).

We see no preferred angle between inner and outer bars, in either the
WIYN or expanded samples.  This agrees with previous results (Buta \&
Crocker 1993; Wozniak et al.\ 1995) and supports models where inner
bars rotate at different speeds than outer bars (Maciejewski \& Sparke
2000).  About half of the inner bars in the WIYN sample are surrounded
by nuclear rings --- dusty, star-forming, or purely stellar.  About
one-third of the WIYN sample inner bars are devoid of strong dust
lanes, while two are found in the midst of substantial
\textit{off-plane} gas.  Structurally, inner bars appear similar to
the outer bars (see Figure~2), with the ``flat'' luminosity profiles
seen in large bars in early-type galaxies (Elmegreen et al.\ 1996b).

Finally, the presence or absence of inner bars has \textit{no}
statistically significant effect on nuclear activity in the WIYN
sample: galaxies with only one bar are as likely to host AGNs as those
with two.

\subsection{Inner disks}

Here, our classification is based purely on orientation: an elliptical
structure (within a large bar) which is aligned to within 10\deg{} of
the galaxy's outer disk is called an inner disk, as long as unsharp
masking does not show it to be, e.g., a nuclear ring instead. 
Clearly, this class will include at least some inner bars with chance
alignments, along with unresolved nuclear rings and possibly flattened
inner bulges.  However, there are several pieces of evidence
indicating that our inner disks form a physically distinct class.

In the WIYN sample, inner disks are as common as inner bars in S0
galaxies, but practically absent in later Hubble types: 35\% of the S0
galaxies had an inner disk, while we found one inner disk in an S0/a
galaxy and none in Sa galaxies.  This contrasts strongly with inner
bars, which are equally common for S0, S0/a, and Sa galaxies.  Inner
disks are also too common for them \textit{all} to be chance
alignments of randomly oriented inner bars.  Finally, inner disks are
systematically larger with respect to their parent bars (median size =
19\% for WIYN sample, 20\% for expanded sample); K-S tests indicate
that inner bars and disks come from different parent populations at a
99.8\% confidence level (96\% for the WIYN sample).  Curiously, inner
disks are not significantly larger than inner bars in absolute size,
or as a fraction of $R_{25}$; the difference is due mostly to the bars
of inner-disk galaxies being systematically smaller (median radius =
3.9 kpc or 0.39 $R_{25}$) than the (outer) bars of double-barred
galaxies (5.7 kpc or 0.52 $R_{25}$).

\begin{figure}
\plotone{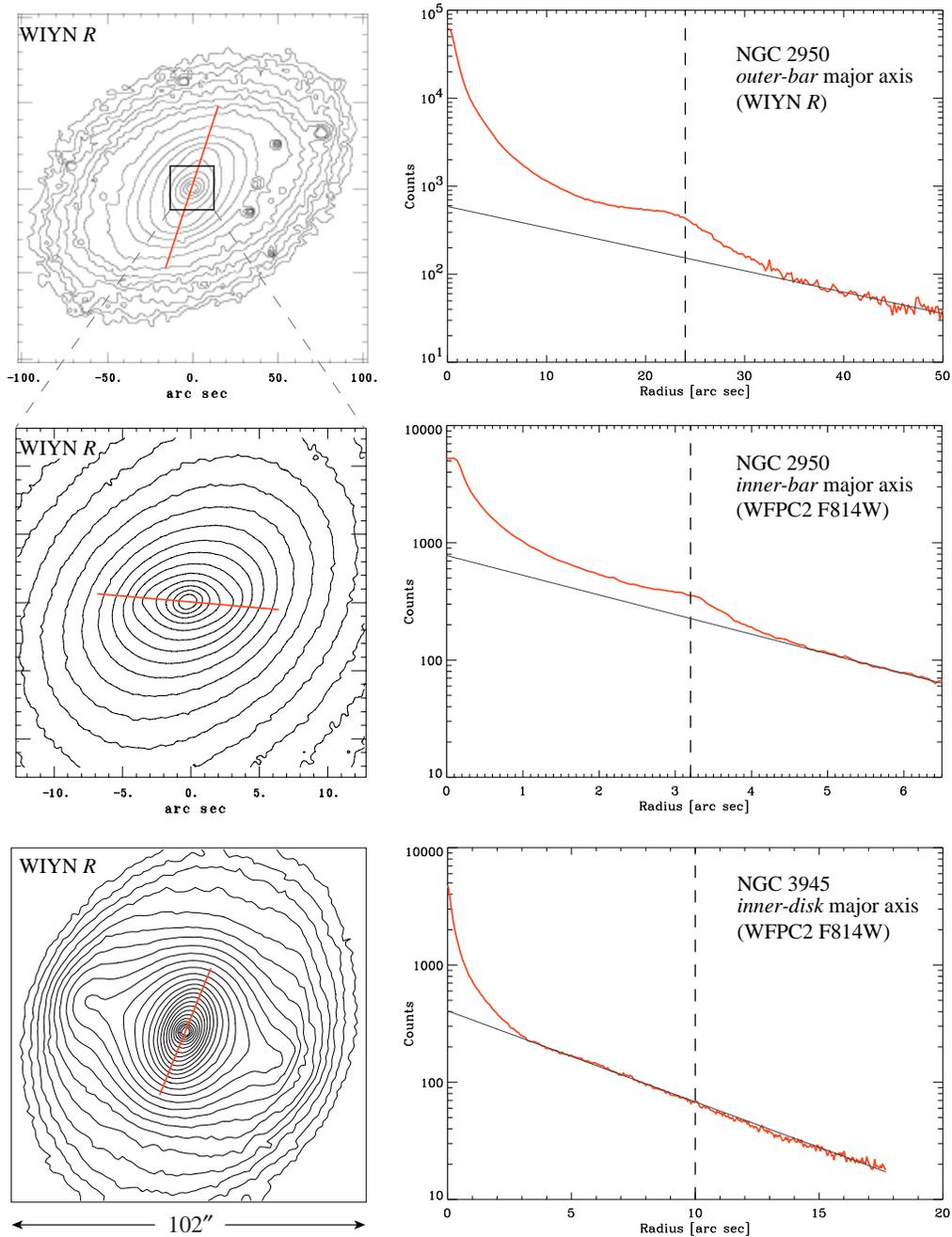}

\caption{Differing luminosity profiles of inner bars and disks.  Top,
major-axis profile (right) of the \textit{outer} bar in double-barred
galaxy NGC 2950; middle, major-axis profile of its \textit{inner} bar
(the innermost points are from saturated pixels); bottom, major-axis
profile of the inner disk in NGC 3945.  In each profile plot, the
vertical dashed line marks the semi-major axis of maximum ellipticity
for the bar or disk, based on ellipse fits; the slanted lines are
exponential fits by eye to the outer parts of each profile.  Note the
similar structure of both inner and outer bar profiles, with a
characteristic break near the bar end; the inner disk is quite
different, and resembles a simple disk + bulge system.}

\end{figure}

At least some of the structures we label inner disks have radial
profiles similar to larger-scale galactic disks, and different from
bars.  Figure~2 shows that the inner disk of NGC 3945 (the largest and
brightest in our sample) differs dramatically in its luminosity
profile from a typical inner bar; it has an exponential profile with a
scale length $\sim 500$ pc, and is distinct from the inner bulge
profile.  Radial luminosity profiles may be a useful tool for
determining the true identity of inner disks.  Kinematic data will be
needed to determine if these inner disks are dynamically cool; this
would also distinguish them from oblate exponential bulges.

\subsection{Nuclear Rings, Spirals, and Off-Plane Gas}

We find nuclear rings in a quarter of the WIYN sample galaxies.  Most
of these rings are dusty and/or star-forming; such rings are
preferentially found in Sa galaxies.  Two of the S0 galaxies have
purely stellar nuclear rings, which may be the remnants of past
circumnuclear starburst episodes (see Erwin, Vega, \& Beckman, this
volume, for more details).  We find nuclear spirals in $\approx 20$\%
of the galaxies; two of these have blue, luminous stellar arms,
indicating recent star formation, but the other six are seen only in
dust absorption.  Finally, 30\% of the S0 galaxies have evidence for
off-plane gas in the central kiloparsec, in addition to misaligned H I
gas at much larger radii.

\end{document}